\documentclass[11pt,a4paper]{article}

\usepackage[margin=2.2cm]{geometry}
\usepackage{lmodern}
\usepackage{amsmath,amssymb}
\usepackage{graphicx}
\DeclareGraphicsExtensions{.pdf,.eps,.png}
\usepackage{booktabs}
\usepackage{multirow}
\usepackage{float}
\usepackage[T1]{fontenc}
\usepackage[utf8]{inputenc}
\usepackage{newunicodechar}
\newunicodechar{ρ}{$\rho$}
\newunicodechar{τ}{$\tau$}
\newunicodechar{η}{$\eta$}
\usepackage{hyperref}
\usepackage[numbers,sort&compress]{natbib}
\usepackage[skip=0.45em plus 0.1em minus 0.05em]{parskip}

\graphicspath{{./}}

\title{Aluminum wide-range EOS sensitivity in Z-pinch dynamic-hohlraum liner--foam impact and outgoing radiation}

\author{%
Cong Wang$^{1,2,3,4,*}$, Zi Li$^{1,2,3}$, Weijie Li$^{3}$, and Xianjue Peng$^{1,3}$\\[0.5em]
\small
$^{1}$Institute of Applied Physics and Computational Mathematics, Beijing 100088, China\\
$^{2}$National Key Laboratory of Computational Physics, Beijing 100088, China\\
$^{3}$Tianfu Institute of Innovative Energy Research, Chengdu, Sichuan 610213, China\\
$^{4}$Center for Applied Physics and Technology, Peking University, Beijing 100871, China\\[0.3em]
$^{*}$Corresponding author. E-mail: wang\_cong@iapcm.ac.cn
}

\date{}

\begin{document}
\maketitle

\begin{abstract}
In a Z-pinch dynamic hohlraum, an aluminum liner collides with a central foam and converts kinetic energy into radiation, and the liner--foam response depends on the aluminum equation of state (EOS). We construct a wide-range aluminum equation-of-state table from density functional theory (DFT) simulations over solid--liquid--plasma regions, denoted DFT-eos. Mixed deterministic--stochastic finite-temperature DFT (mDFT) gives the electron thermal contribution; quantum molecular dynamics (QMD) with classical calibration gives the ion part. We perform one-dimensional cylindrical radiation magnetohydrodynamics (MHD) simulations at a peak current of 50~MA and rise times of 220--300~ns on a single-layer light foam load. Opacity and electrical conductivity are fixed; only the liner Al EOS is changed between DFT-eos and the quotidian EOS (QEOS). Impact time and peak liner kinetic energy change little between the two tables. Larger differences appear in flight-segment density, the contact state at impact, and the time history of outgoing radiation power at the outer interface.
\end{abstract}



\section{Introduction}

In a dynamic hohlraum, an imploding shell (wire array or liner) collides with a central foam and converts kinetic energy into radiation that drives a capsule \cite{Nash1999,Sanford2002,Rochau2008}. MagLIF experiments already show high-current liner implosions \cite{Slutz2010,Gomez2020,HarveyThompson2024,Weis2025,Shipley2024}. At ~50 MA, dynamic-hohlraum coupling to double-shell targets is still studied mainly in simulations \cite{Chu2022,Li2025DS}. After impact, a radiative shock propagates through the foam. The radiation temperature and outgoing radiation power at the outer interface are set by radiative transport and depend on load dynamics and material properties \cite{Rochau2008,Wu2018}. Wire-array--foam coupling, radiation uniformity, and dynamic-hohlraum radiation sources have been studied in theory and experiment \cite{Nash1999,Sanford2002,Rochau2007,Rochau2008,Ding2016,Wu2018,Huang2019,Meng2017,Ye2020,Yi2023}. To quantify energy conversion in a dynamic hohlraum, liner state evolution before and after impact, foam heating, and outer-interface radiation must be modeled together.

Dynamic-hohlraum designs that target higher peak current and longer rise times have received attention, including drivers near 50 MA with rise times of hundreds of nanoseconds \cite{Cheng2023,DelongXiao2021,Chu2022,Li2025DS}. Aluminum liners are light and implode with relatively clean dynamics; paired with a low-density foam converter they form a standard geometry for kinetic-to-radiation conversion \cite{Sanford2002,Ye2020}. A liner starts near ambient solid, expands through melting and vaporization, and heats strongly on impact, so a wide-range EOS spanning solid--liquid--plasma states is required.

First-principles calculations and average-atom models cover warm-dense and hot-dense matter along several complementary routes. Average-atom schemes use a spherically averaged atomic potential and return properties quickly over a wide (ρ, T) range \cite{Liberman1982}. Path-integral Monte Carlo (PIMC), combined with DFT-MD, provides a warm-dense benchmark for aluminum and resolves atomic shell structure in the EOS \cite{Driver2018,Militzer2021}. Kohn--Sham QMD fits molten and warm-dense liquid states and portions of the shock path \cite{Sjostrom2016,Kumar2023}. Orbital-free molecular dynamics (OFMD) extends ion dynamics to higher temperature \cite{Lambert2006}. Each method has a practical limit: average-atom models lack explicit ion environments and disorder; PIMC cost rises sharply at low temperature; QMD orbital counts grow with temperature; OFMD sacrifices shell resolution through orbital-free kinetics. Mixed stochastic--deterministic DFT (mDFT) provides the electron thermal EOS from finite-temperature DFT and retains shell structure and partial ionization where full QMD is too expensive at high temperature \cite{Chen2024,Chen2025}.

Most benchmarks supply state points or partial tables rather than one table from cold solid through two-phase regions to hot dense plasma. Wide-range radiation MHD calculations for aluminum therefore still use SESAME \cite{Lyon1992} or QEOS \cite{More1988}. The electron thermal part and principal Hugoniot of warm-dense Al nevertheless depend on the electron model \cite{main_pop,Chen2025}. X-ray Thomson scattering on shock-compressed Al shows that a uniform electron-gas model does not capture plasmon dispersion without ion disorder and correlations at first-principles level \cite{Bespalov2026,Gawne2024}. Shock melting and Hugoniot temperatures in Al also face new experimental constraints \cite{Jin2025,Hartsfield2026}.

In this work, we build a wide-range Al liner EOS table, DFT-eos (mDFT electrons and QMD-calibrated ions; assembly in \S{}II), for this implosion problem. We apply DFT-eos and QEOS in a one-dimensional cylindrical radiation MHD code. We use a 50 MA peak current, rise times of 220--300 ns, and a single-layer light foam load; the drive, foam properties, and transport coefficients are held fixed.

The remainder of this paper is organized as follows: \S{}II describes the construction of DFT-eos, phase-region matching, and comparison with QEOS; \S{}III gives the load, drive, and diagnostics; \S{}IV compares liner flight, contact at impact, energy partition, and radiation output; \S{}V discusses the mechanisms; \S{}VI summarizes the conclusions.

\section{Aluminum equation of state}

The liquid--plasma branch of DFT-eos follows the free-energy decomposition in \cite{main_pop}; here we complete the solid branch in the same framework and assemble the full solid--liquid--plasma--vapor table used in \S{}III--IV. The dilute vapor phase is modeled as a monatomic ideal gas, and phase boundaries connect to the condensed branches through Maxwell constructions at equal G and P \cite{Faik2018}. For comparison, QEOS \cite{More1988} is generated with FEOS \cite{Faik2018}; the foam CH EOS is likewise held at QEOS. Multigroup opacity is taken from Los Alamos Planck/Rosseland tables (TOPS/OPLIB) \cite{Magee1995} and electrical conductivity from the Lee--More model \cite{LeeMore1984}; only the Al liner EOS is varied.

\subsubsection*{A. Solid region}

In the solid region we retain only the fcc phase. Other high-pressure solid phases are not included, because the liner in our MHD loads never reaches the pressures and temperatures at which they would appear. The fcc solid free energy is

\[F_\mathrm{s}(V,T)=E_c(V)+F_\mathrm{ion}^\mathrm{(Debye)}(V,T)+F_e(\rho,T),\]

where \(E_c\) follows a cohesive-form Vinet cold curve, \(F_\mathrm{ion}^\mathrm{(Debye)}\) is a single-Debye ion thermal term, and \(F_e(\rho,T)\) is taken from the same mDFT electron thermal table as the liquid--plasma branch (\S{}II.B). The cold energy is

\[E_c(V)=\varphi_0+\frac{4V_0 B_0}{(B_0'-1)^2} \Big[1-(1+\chi)\,e^{-\chi}\Big],\]

with \(\chi=\tfrac{3}{2}(B_0'-1)[(V/V_0)^{1/3}-1]\). In the separated-atom limit \(V\to\infty\), \(E_c\to0\) and \(E_c(V_0)=\varphi_0\). The Debye temperature is

\[\theta(V)=\theta_0\,(V_0/V)^{\gamma_0},\]

and the ion thermal free energy is

\[F_\mathrm{ion}^\mathrm{(Debye)}(V,T)=\frac{9}{8}k_B\theta(V)+k_B T\left[3\ln(1-e^{-\theta/T})-D(\theta/T)\right],\]

where \(D(y)=(3/y^3)\int_0^y t^3/(e^t-1)\,dt\) is the Debye function. Parameters are \(\rho_0=2.70\,\mathrm{g\,cm^{-3}}\) (\(V_0\simeq16.59\,\mathrm{\AA^3/atom}\)), \(B_0=75\,\mathrm{GPa}\), \(B_0'=4.6\), \(\varphi_0=-3.36\,\mathrm{eV/atom}\), \(\theta_0=420\,\mathrm{K}\), and \(\gamma_0=2.1\). The wide-range table uses the dilute monatomic vapor as the energy zero (\(E\to0\)), with binding at \(\rho_0\) relative to separated atoms.

\subsubsection*{B. Liquid--plasma region}

The liquid--plasma branch uses the decomposition \cite{main_pop}

\[F(\rho,T)=F_c(\rho)+F_\mathrm{ion}(\rho,T)+F_e(\rho,T).\]

The cold contribution \(F_c\) is obtained from Kohn--Sham QMD total energies in the warm-dense region, extrapolated isothermally to \(T\to0\), and fitted with a Vinet--Rose form. The fit gives \(\rho_\mathrm{ref}=2.297\,\mathrm{g\,cm^{-3}}\), \(B_0=43.25\,\mathrm{GPa}\), and \(B_0'=4.955\); the cold pressure is \(P_c=-(\partial F_c/\partial V)_T\).

The ion thermal part combines a Debye term with a finite-cell correction,

\[F_\mathrm{ion}=F_\mathrm{ion}^\mathrm{(Debye)}+F_\mathrm{cell},\]

where \(F_\mathrm{cell}=-k_B T\ln Q\) is a soft-sphere, finite-cell correction with \(Q\) of the form \(\mathrm{erf}\sqrt{T^\ast/T}\), in the same spirit as the Johnson-type ion thermal model for liquid Al \cite{Sjostrom2016}. The Debye temperature \(\theta(V)\) follows the piecewise liquid Gr\"{u}neisen parameter through \(\mathrm{d}\ln\theta=\gamma\,\mathrm{d}\ln\rho\) \cite{Sjostrom2016}; the ionic thermal pressure is least-squares fitted to QMD data after subtracting the cold and electronic contributions \cite{main_pop}.

The electron thermal contribution \(F_e\) is built from mDFT calculations, tabulated on a \((\rho,T)\) grid and used with bilinear interpolation \cite{main_pop}. On the same grid, Thomas--Fermi--Dirac (TFD) and ionization-equilibrium (IONEQ) models deviate from mDFT in pressure by about \(+59\%\) and \(-31\%\) near \(\rho=12\,\mathrm{g\,cm^{-3}}\) and \(T=10\,\mathrm{eV}\). On the principal Hugoniot, mDFT, TFD, and IONEQ agree for \(\rho\lesssim8\,\mathrm{g\,cm^{-3}}\) but diverge at higher compression \cite{main_pop}. DFT-eos uses the mDFT electron thermal table as the electron branch throughout the liquid--plasma region.

\subsubsection*{C. Principal Hugoniot from solid to plasma}

The principal Hugoniot is obtained from the Rankine--Hugoniot relations on the full DFT-eos table, with reference state \(\rho_0=2.70\,\mathrm{g\,cm^{-3}}\) and \(T_0=300\,\mathrm{K}\). Fig. 1 shows the Hugoniot from this table: on the \(P\)--\(T\) plane, the curve rises along the solid branch, intersects the Maxwell melt line \(G_\mathrm{s}=G_\ell\) near shock melting (about \(130\,\mathrm{GPa}\) and \(4800\,\mathrm{K}\)), and enters the liquid branch through the melt-line window. The melt line follows coexistence with \(G_\mathrm{s}=G_\ell\) and \(P_\mathrm{s}=P_\ell\), in good agreement with \cite{Vocadlo2002}. At higher compression, checks against first-principles reference data and wide-range tables are documented in \cite{main_pop} and are omitted here.

\subsubsection*{D. Comparison with QEOS}

QEOS \cite{More1988} provides a wide-range Al table from a modified Thomas--Fermi electron model and a multiphase ion thermal model; the QEOS used here is generated with FEOS \cite{Faik2018}. The main differences from DFT-eos are as follows.

(1) Near ambient solid conditions, DFT-eos includes the cohesive energy \(\varphi_0\) and a Debye solid branch and gives \(P(\rho_0,300\,\mathrm{K})\approx0\). QEOS is a wide-range multiphase fit with a different solid branch and cohesion depth. The liner initial state and early expansion path depend on solid cohesion and the solid branch.

(2) In the warm-dense liquid--plasma region, DFT-eos uses the mDFT electron branch, whereas QEOS uses its modified Thomas--Fermi electron branch. At the same \((\rho,T)\), the QEOS electron pressure is usually higher than mDFT because the Thomas--Fermi treatment does not resolve shell structure and does not adequately describe coupling and partial excitation \cite{main_pop}. This matches the isobaric flight comparison in which DFT-eos gives lower pressure than QEOS (Sec. IV.1, Fig. 1(b)).

(3) In low- to moderate-density expansion and two-phase regions, DFT-eos defines solid--liquid and solid--vapor/liquid--vapor coexistence bands explicitly through Maxwell constructions at equal \(G\), joined to the vapor branch. Along the liner expansion path, QEOS tends to keep a lower density in the low-density segment; Sec.~IV quantifies the flight-segment contrast (Table 2).

(4) On the strongly compressed Hugoniot, see Sec. II.C. As compression increases, differences in the electron model separate the DFT-eos (mDFT) path from TF/IONEQ and QEOS; details are given in \cite{main_pop}.

On the liner path to impact, peak inner Al densities are mostly \(\sim1\)--\(2\,\mathrm{g\,cm^{-3}}\) with electron temperatures of hundreds of eV. The load therefore samples warm-dense and low- to moderate-density regions rather than the extreme-compression regime emphasized in \cite{main_pop}.

\begin{figure}[htbp]
\centering
\includegraphics[width=\linewidth]{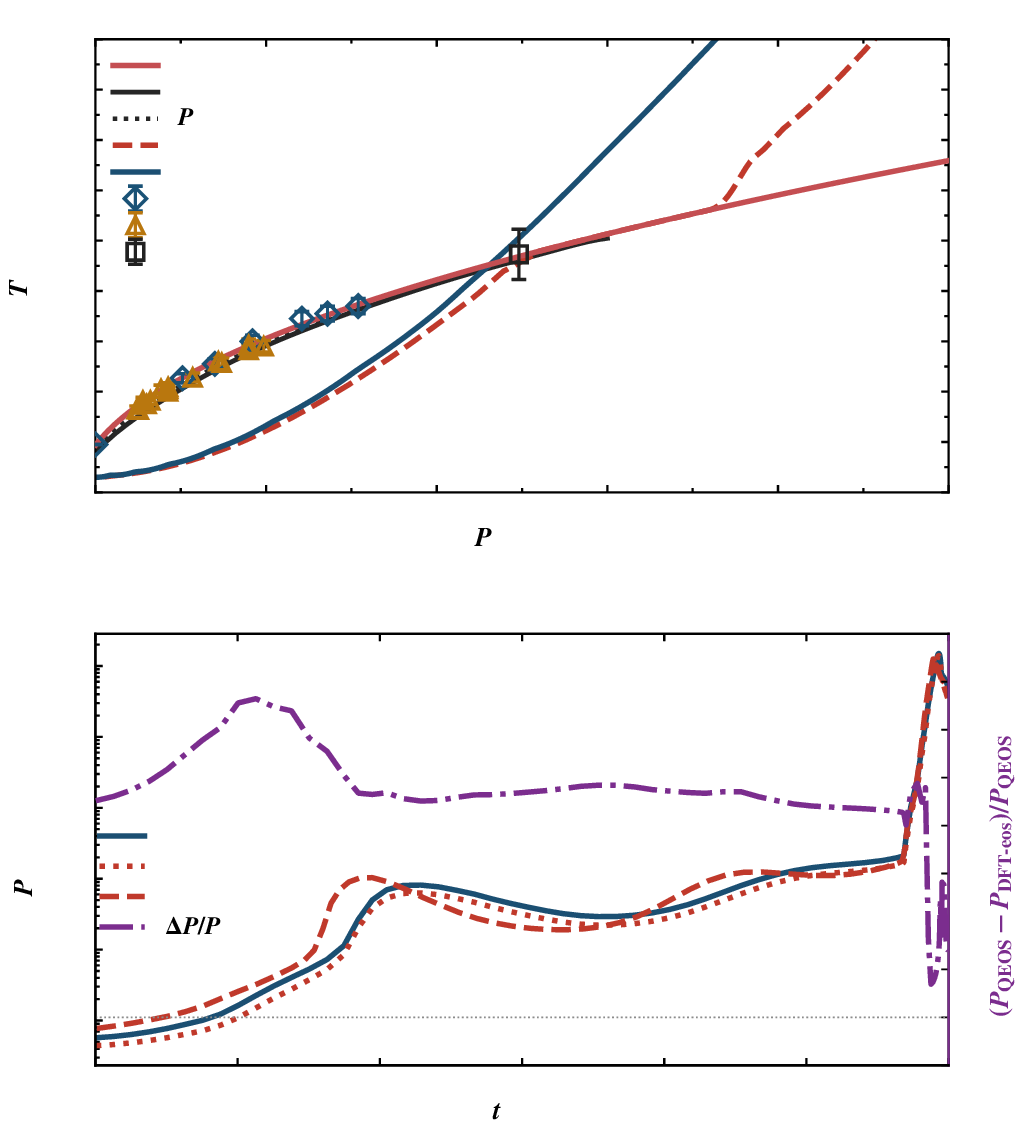}
\caption{Aluminum EOS checks. (a) \(P\)--\(T\): red curve is the DFT-eos Maxwell melt from \(G_\mathrm{s}=G_\ell\). Black solid and dotted curves: GGA and pressure-corrected melts from \cite{Vocadlo2002}, Fig. 1. Open diamonds: \cite{Boehler1997}. Open triangles: \cite{Hanstrom2000}. Open square: \cite{Shaner1984} shock-melt point. Blue solid = QEOS Hugoniot, red dashed = DFT-eos Hugoniot, shown to about \(250\,\mathrm{GPa}\). (b) Flight window at \(\tau=260\,\mathrm{ns}\): table \(P\) along QEOS and DFT-eos trajectories; right axis \((P_{\mathrm{QEOS}}-P_{\mathrm{DFT\text{-}eos}})/P_{\mathrm{QEOS}}\) on the QEOS path (sign opposite to the paper-wide relative difference \((\mathrm{DFT\text{-}eos}-\mathrm{QEOS})/|\mathrm{QEOS}|\), so a positive value means QEOS is stiffer).}
\label{fig:1}
\end{figure}
\section{Numerical methods}

Calculations use a one-dimensional Lagrangian cylindrical radiation MHD code. Implicit hydrodynamics, multigroup diffusive radiation transport, and resistive MHD with a prescribed current waveform are solved on tabulated EOS and opacity data, with ohmic dissipation from the drive current included in the energy closure. The peak drive current is \(I_\mathrm{max}=50\,\mathrm{MA}\) with a \(\sin^4\) waveform; rise times are \(\tau=220\), 260, and 300 ns, and the current peaks at \(t\approx\tau\). The Al liner inner radius is \(R_\mathrm{in}=4.0\,\mathrm{cm}\) and the line mass is \(m=20.0\,\mathrm{mg/cm}\) (cold thickness \(\sim2.95\,\mu\mathrm{m}\)), discretized with 100 radial zones. The foam is a single-layer CH shell with outer radius \(R=0.50\,\mathrm{cm}\) and initial density \(\rho_0=0.015\,\mathrm{g/cm^3}\), discretized with 120 radial zones. The vacuum gap between liner and foam is 3.50 cm. The load cross section and current waveform are shown in Fig. 2; main parameters are in Table 1.

\begin{table}[htbp]
\centering
\small
\caption{Baseline load and drive parameters}
\label{tab:Table_1}
\begin{tabular*}{\textwidth}{@{\extracolsep{\fill}}l l}
\hline
Parameter & Value \\
\hline
\(I_\mathrm{max}\) & 50 MA \\
\(\tau\) & 220 / 260 / 300 ns \\
\(R_\mathrm{in}\) & 4.0 cm \\
\(m\) & 20.0 mg/cm \\
Al radial zones & 100 \\
Foam & Single-layer CH, \(R=0.50\,\mathrm{cm}\), \(\rho_0=0.015\,\mathrm{g/cm^3}\) \\
Foam radial zones & 120 \\
Vacuum & 3.50 cm \\
\hline
\end{tabular*}
\end{table}

\begin{figure}[htbp]
\centering
\includegraphics[width=\linewidth]{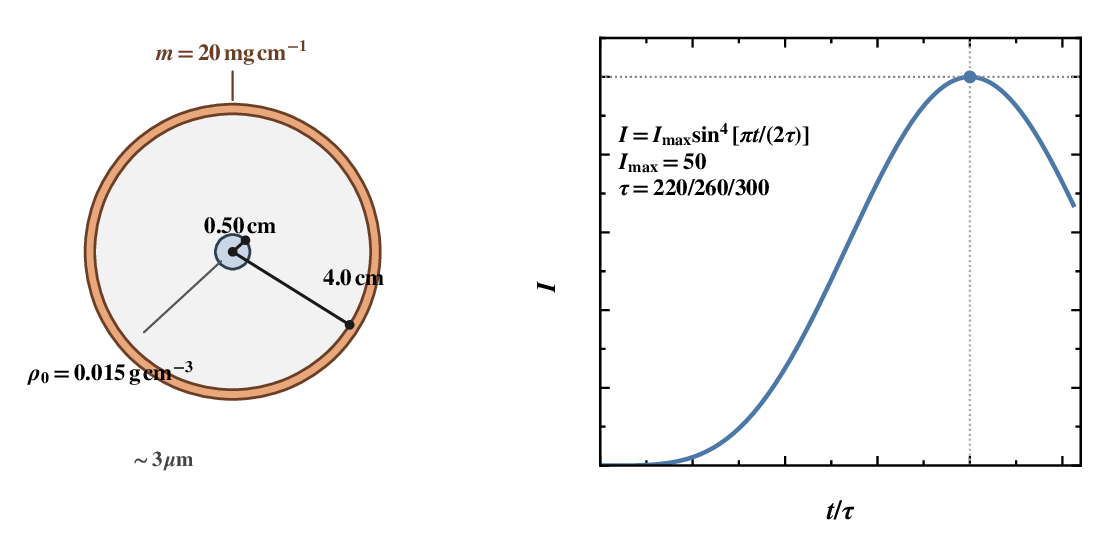}
\caption{Load geometry and drive current. (a) End-on radial cross section: central CH foam outer radius \(R=0.50\,\mathrm{cm}\), Al liner inner radius \(R_\mathrm{in}=4.0\,\mathrm{cm}\), vacuum between them. (b) Normalized current \(I/I_\mathrm{max}\) with \(\sin^4\) waveform, \(I_\mathrm{max}=50\,\mathrm{MA}\); \(\tau=220\), 260, and 300 ns share the same waveform shape.}
\label{fig:2}
\end{figure}

Liner kinetic energy \(E_k\) is taken as the peak before impact. Foam thermal internal energy \(E_\mathrm{th}\) and peak Al inner-state quantities are maxima within a 15 ns window after impact. Thermal conversion efficiency \(\eta=E_{\mathrm{th,foam,peak}}/E_{k,\mathrm{peak}}\), reported in percent, uses the peak foam thermal internal energy within the post-impact 15 ns window and excludes foam kinetic energy. Al inner density and electron temperature are mass-weighted averages over the inner one-fifth of the liner radial zones. Radiation temperature \(T_r\) and outgoing radiation power \(S^+\) are taken at the outer foam interface; \(S^+\) is in TW. The full width at half maximum (FWHM) of \(S^+(t)\) is the difference between the first and last times above half peak within the post-impact 15 ns window. Results distinguish impact time and post-impact peak time. Relative differences are \((\mathrm{DFT\text{-}eos}-\mathrm{QEOS})/|\mathrm{QEOS}|\times100\%\).

\section{Results}

In this dynamic-hohlraum-style load, a rising current drives an aluminum liner radially inward in cylindrical geometry. Magnetic pressure implodes the Al liner inward across a vacuum gap until the liner contacts the low-density CH foam shell. During the long interval before contact, expansion and ohmic heating leave the liner in a warm, rarefied state with average density far below solid aluminum, while the inner radius continues to close. At impact, liner kinetic energy is converted to shock heating and thermal energy in the foam and in the inner liner layers; multigroup radiation transport then carries energy outward through the foam, where the escaping radiation is diagnosed. Under the fixed drive and foam configuration in Sec.~III, we compare DFT-eos and QEOS while varying only the Al liner EOS table.

\subsection{Macroscopic implosion and flight segment}

\begin{table}[htbp]
\centering
\caption{Peak liner kinetic energy, flight-segment density at hit-50 ns, and impact velocity. Units: \(E_{k,\mathrm{peak}}\) [MJ/cm]; \(\bar\rho\) [g/cm$^{3}$]; \(u_\mathrm{hit}\) [cm/$\mu$s]. Column headers QEOS and DFT-eos are the two Al liner EOS tables; \(d\%=(\mathrm{DFT\text{-}eos}-\mathrm{QEOS})/|\mathrm{QEOS}|\times100\%\). \(\bar\rho\) is the liner mass-weighted density at \(t=t_\mathrm{hit}-50\,\mathrm{ns}\); \(u_\mathrm{hit}\) is the inward velocity of the liner inner interface at impact.}
\label{tab:Table_2}
\begin{tabular*}{\textwidth}{@{\extracolsep{\fill}}lccccccccc}
\toprule
 & \multicolumn{3}{c}{\(E_{k,\mathrm{peak}}\)} & \multicolumn{3}{c}{\(\bar\rho\)} & \multicolumn{3}{c}{\(u_\mathrm{hit}\)} \\
\cmidrule(lr){2-4}\cmidrule(lr){5-7}\cmidrule(lr){8-10}
\(\tau\) [ns] & QEOS & DFT-eos & $d\%$ & QEOS & DFT-eos & $d\%$ & QEOS & DFT-eos & $d\%$ \\
\midrule
220 & 2.431 & 2.524 & +3.8 & 0.00836 & 0.0138 & +65.0 & 25.92 & 32.99 & +27.3 \\
260 & 3.327 & 3.460 & +4.0 & 0.00859 & 0.0129 & +49.9 & 35.96 & 40.92 & +13.8 \\
300 & 3.864 & 3.984 & +3.1 & 0.00789 & 0.0117 & +48.0 & 40.43 & 48.60 & +20.2 \\
\bottomrule
\end{tabular*}
\end{table}

Over \(\tau=220\)--300 ns, \(t_\mathrm{hit}\) from QEOS is 296.7, 321.6, and 347.9 ns and from DFT-eos is 297.5, 321.6, and 347.6 ns (relative differences \(\lesssim0.3\%\)). Peak liner kinetic energy varies by \(\lesssim4\%\); DFT-eos gives slightly higher \(E_{k,\mathrm{peak}}\) (Table 2). Fig. 3(a--b) (\(\tau=260\,\mathrm{ns}\)) shows nearly overlapping \(R_\mathrm{in}(t)\), \(R_\mathrm{out}(t)\), and liner \(E_k(t)\). By contrast, flight-segment \(\bar\rho\) and \(u_\mathrm{hit}\) differ by tens of percent, with DFT-eos giving higher \(\bar\rho\) at hit-50 ns and higher \(u_\mathrm{hit}\) on all three rise times (Table 2). At hit-50 ns, \(\bar\rho\ll0.1\,\mathrm{g/cm^3}\) on both sides. Fig. 3(c) traces \(\langle\rho\rangle(t)\); DFT-eos stays above QEOS throughout the flight window. Fig. 1(b) gives table pressure along the two trajectories in the \(\tau=260\,\mathrm{ns}\) flight window; the right axis \((P_\mathrm{QEOS}-P_\mathrm{DFT\text{-}eos})/P_\mathrm{QEOS}\) uses the opposite sign convention to the paper-wide \(d\%\), so a positive value means QEOS is stiffer at the same state.

\begin{figure}[htbp]
\centering
\includegraphics[width=\linewidth]{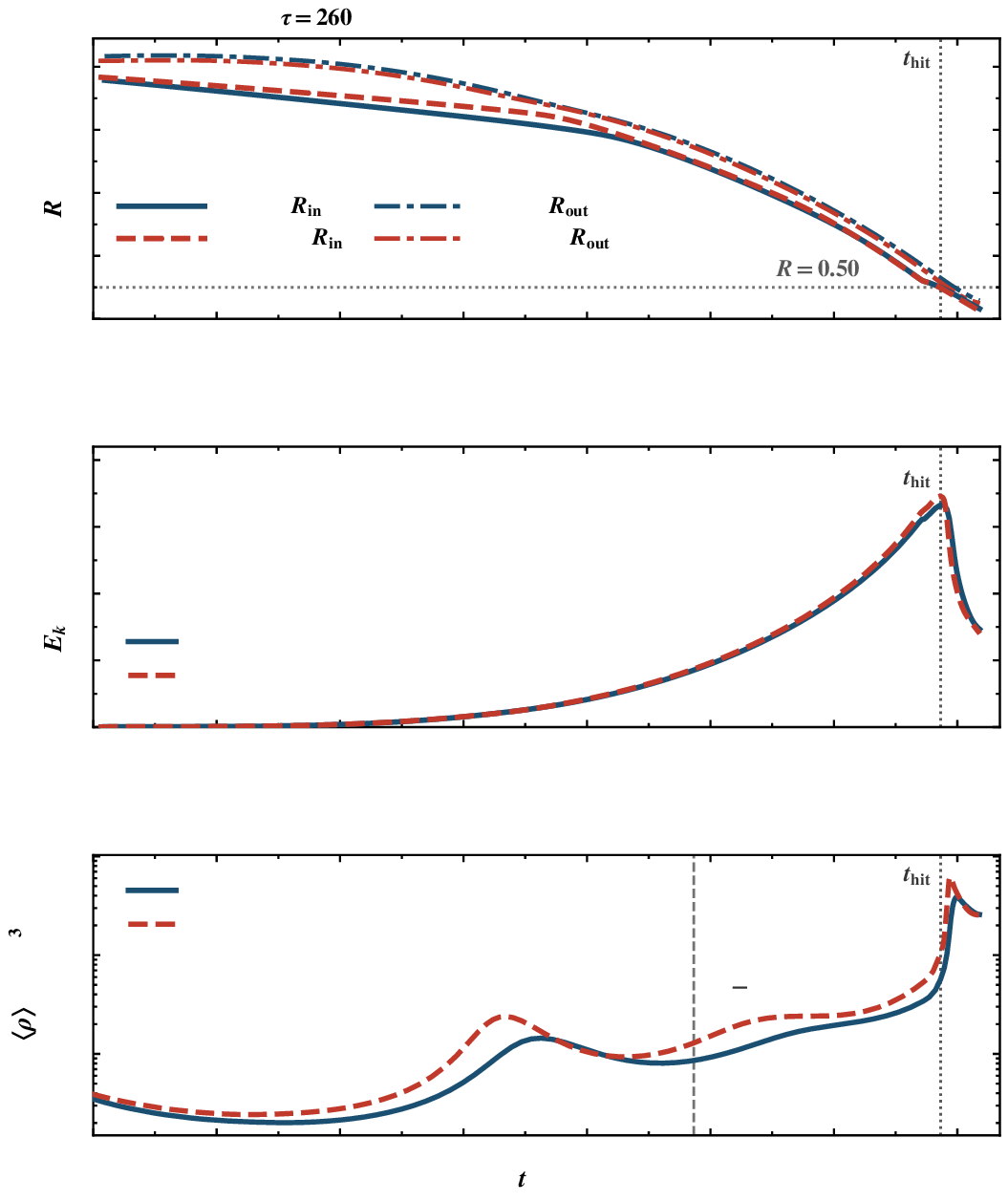}
\caption{Macroscopic implosion and flight rarefaction at \(\tau=260\,\mathrm{ns}\). (a) Liner \(R_\mathrm{in}(t)\) (solid/dashed) and \(R_\mathrm{out}(t)\) (dash--dot); CH foam outer radius as marked. (b) Liner \(E_k(t)\). (c) Liner mass-weighted \(\langle\rho\rangle(t)\); vertical guides mark hit-50 ns (c) and \(t_\mathrm{hit}\) (all panels). Other rise times are in Table 2.}
\label{fig:3}
\end{figure}

\subsection{Impact contact state and energy partition}

\begin{table}[htbp]
\centering
\caption{Inner Al contact state and outer-interface \(T_r\). Units: \(\rho\) [g/cm$^{3}$]; \(T_e\), \(T_r\) [eV]. (a) At \(t_\mathrm{hit}\). (b) Maxima within 15 ns after impact (Fig.~4). Column headers QEOS and DFT-eos are the two Al liner EOS tables; \(d\%=(\mathrm{DFT\text{-}eos}-\mathrm{QEOS})/|\mathrm{QEOS}|\times100\%\). \(\rho_\mathrm{in}\) and \(T_{e,\mathrm{in}}\) are mass-weighted over the inner one-fifth of liner radial zones; \(T_r\) is at the outer foam interface.}
\label{tab:Table_3}
\vspace{0.35em}
\renewcommand{\arraystretch}{1.15}
\textbf{(a)} At \(t_\mathrm{hit}\)\par\vspace{0.25em}
\begin{tabular*}{\textwidth}{@{\extracolsep{\fill}}lccccccccc}
\toprule
 & \multicolumn{3}{c}{\(\rho_\mathrm{in}\)} & \multicolumn{3}{c}{\(T_{e,\mathrm{in}}\)} & \multicolumn{3}{c}{\(T_r\)} \\
\cmidrule(lr){2-4}\cmidrule(lr){5-7}\cmidrule(lr){8-10}
\(\tau\) [ns] & QEOS & DFT-eos & $d\%$ & QEOS & DFT-eos & $d\%$ & QEOS & DFT-eos & $d\%$ \\
\midrule
220 & 0.04901 & 0.118 & +140.1 & 135.6 & 131.7 & $-$2.9 & 110.9 & 123.4 & +11.2 \\
260 & 0.09149 & 0.155 & +69.6 & 192.6 & 150.1 & $-$22.1 & 122.6 & 137.4 & +12.1 \\
300 & 0.1722 & 0.178 & +3.3 & 336.1 & 160.1 & $-$52.4 & 131.9 & 145.8 & +10.5 \\
\bottomrule
\end{tabular*}
\vspace{0.85em}
\textbf{(b)} Peak within 15 ns after impact\par\vspace{0.25em}
\begin{tabular*}{\textwidth}{@{\extracolsep{\fill}}lccccccccc}
\toprule
 & \multicolumn{3}{c}{\(\rho_\mathrm{in}\)} & \multicolumn{3}{c}{\(T_{e,\mathrm{in}}\)} & \multicolumn{3}{c}{\(T_r\)} \\
\cmidrule(lr){2-4}\cmidrule(lr){5-7}\cmidrule(lr){8-10}
\(\tau\) [ns] & QEOS & DFT-eos & $d\%$ & QEOS & DFT-eos & $d\%$ & QEOS & DFT-eos & $d\%$ \\
\midrule
220 & 3.001 & 3.296 & +9.8 & 381.6 & 384.5 & +0.8 & 395.8 & 400.3 & +1.1 \\
260 & 4.177 & 4.349 & +4.1 & 524.8 & 512.2 & $-$2.4 & 546.9 & 532.8 & $-$2.6 \\
300 & 4.754 & 5.257 & +10.6 & 610.3 & 584.6 & $-$4.2 & 628.0 & 602.4 & $-$4.1 \\
\bottomrule
\end{tabular*}
\end{table}

At \(t_\mathrm{hit}\), DFT-eos gives higher \(\rho_\mathrm{in}\) while \(T_{e,\mathrm{in}}\) can be lower on longer rise times; outer-interface \(T_r\) is higher with DFT-eos by roughly 10\% on all \(\tau\) (Table 3(a)). Within the post-impact window, relative differences in \(\rho_\mathrm{in}\), mass-weighted \(T_{e,\mathrm{in}}\), and \(T_r\) shrink to the few-percent level (Table 3(b)); grid peak \(T_{e,\mathrm{max}}\) can still differ by \(\sim30\)--\(34\%\) at \(\tau=260\) and 300 ns even when the Fig.~4 contact variables nearly agree. Using post-impact peaks alone therefore understates impact-time contact differences. Fig. 4 (\(\tau=260\,\mathrm{ns}\)) shows the impact-window evolution.

\begin{figure}[htbp]
\centering
\includegraphics[width=\linewidth]{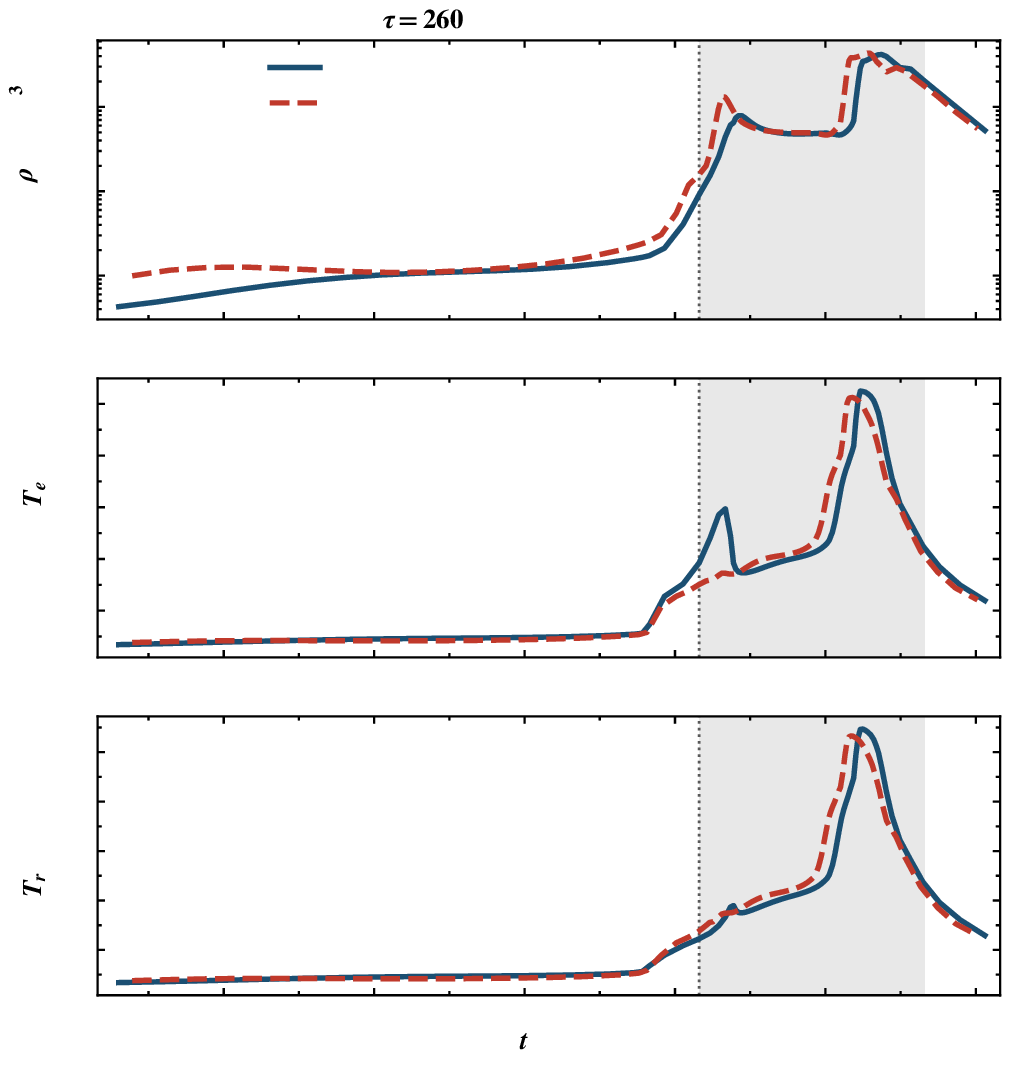}
\caption{Contact state in the impact window (\(\tau=260\,\mathrm{ns}\)). (a) Inner Al \(\rho_\mathrm{in}\); (b) inner Al \(T_{e,\mathrm{in}}\); (c) \(T_r\) at the outer foam interface. Gray band from \(t_\mathrm{hit}\) to 15 ns after impact. Blue solid = QEOS, red dashed = DFT-eos. Results for \(\tau=220\) and 300 ns are in Table 3.}
\label{fig:4}
\end{figure}

Peak Al and foam thermal energy within the same 15 ns window differ by mostly $\lesssim$10\% between EOS tables; \(\eta\) (Sec.~III) is 34.2--39.9\% (QEOS) and 33.5--36.1\% (DFT-eos) over \(\tau=220\)--300 ns, and DFT-eos does not give higher \(\eta\). Peak CH kinetic energy is comparable to peak foam \(E_\mathrm{th}\) (ratio \(\sim97\)--120\%) and is excluded from \(\eta\). Fig. 5(a--b) (\(\tau=260\,\mathrm{ns}\)) shows energy moving from liner kinetic energy into foam thermal and kinetic forms after impact.

\subsection{Outgoing radiation power \(S^+\)}

\begin{table}[htbp]
\centering
\caption{Outgoing \(S^+\) at \(t_\mathrm{hit}\), peak \(S^+\), and pulse FWHM. Units: \(S^+\) [TW]; FWHM [ns]. Column headers QEOS and DFT-eos are the two Al liner EOS tables; \(d\%=(\mathrm{DFT\text{-}eos}-\mathrm{QEOS})/|\mathrm{QEOS}|\times100\%\). Peak \(S^+\) and FWHM are maxima within the post-impact 15 ns window. At \(\tau=260\,\mathrm{ns}\), \(S^+\) peaks at \(t_\mathrm{hit}+10.71\,\mathrm{ns}\) (QEOS) and \(t_\mathrm{hit}+10.13\,\mathrm{ns}\) (DFT-eos).}
\label{tab:Table_4}
\begin{tabular*}{\textwidth}{@{\extracolsep{\fill}}lccccccccc}
\toprule
 & \multicolumn{3}{c}{\(S^+\) at \(t_\mathrm{hit}\)} & \multicolumn{3}{c}{\(S^+_\mathrm{peak}\)} & \multicolumn{3}{c}{FWHM} \\
\cmidrule(lr){2-4}\cmidrule(lr){5-7}\cmidrule(lr){8-10}
\(\tau\) [ns] & QEOS & DFT-eos & $d\%$ & QEOS & DFT-eos & $d\%$ & QEOS & DFT-eos & $d\%$ \\
\midrule
220 & 47.4 & 70.8 & +49.6 & 993.8 & 1115 & +12.2 & 2.98 & 3.54 & +18.8 \\
260 & 70.9 & 110.4 & +55.7 & 2946 & 2825 & $-$4.1 & 2.39 & 3.10 & +29.7 \\
300 & 90.8 & 140.5 & +54.7 & 4414 & 4119 & $-$6.7 & 2.09 & 2.71 & +29.7 \\
\bottomrule
\end{tabular*}
\end{table}

At \(t_\mathrm{hit}\), \(S^+\) differs by roughly 50--56\% with DFT-eos higher (Table 4). Peak \(S^+\) within 15 ns after impact agrees much more closely and can fall below QEOS at \(\tau=260\) and 300 ns, while FWHM stays 20--30\% wider with DFT-eos on all rise times. Fig. 5(c) shows higher \(S^+\) at impact but a slightly lower peak and a broader pulse with DFT-eos---peak \(S^+\) alone is a poor proxy for impact-time radiation.

\begin{figure}[htbp]
\centering
\includegraphics[width=\linewidth]{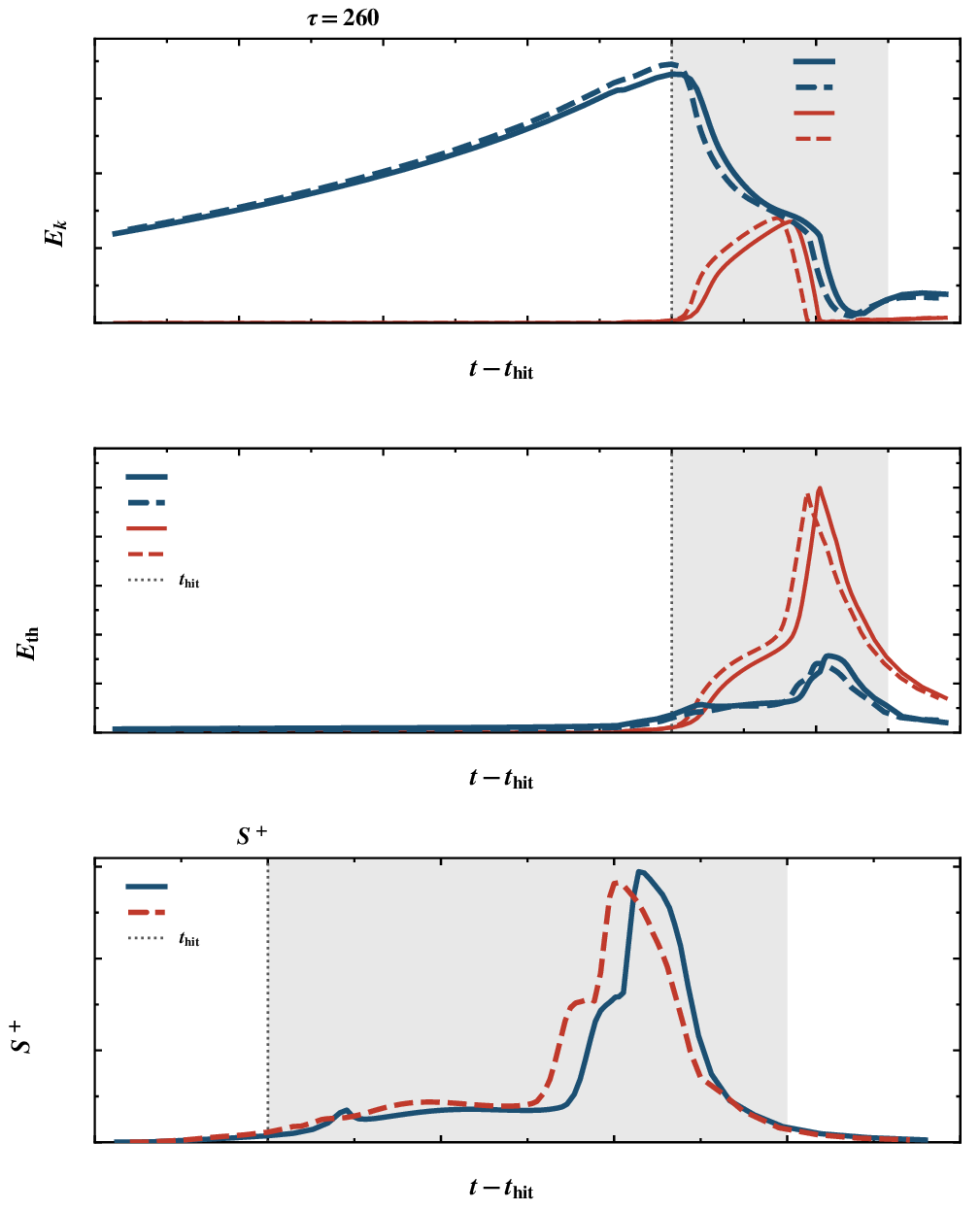}
\caption{Energy partition and outgoing radiation at \(\tau=260\,\mathrm{ns}\). (a) Al and CH kinetic energy \(E_k\); (b) Al and CH thermal internal energy \(E_\mathrm{th}\); (c) outgoing \(S^+\) at the foam outer interface. Horizontal axis: \(t-t_\mathrm{hit}\). Same material, same color in (a--b); EOS by line style (blue solid = QEOS, red dashed = DFT-eos). Gray band: 15 ns after impact. Peak \(S^+\), \(S^+\) at \(t_\mathrm{hit}\), and FWHM are in Table~4.}
\label{fig:5}
\end{figure}

\section{Discussion}

The origin is thermodynamic: the two tables treat Al electron free energy differently. QEOS uses a Thomas--Fermi (TF)--type average model in which ionization varies smoothly with \(\rho\) and \(T\), without explicit shell structure or partial excitation; at the same \((\rho,T)\), \(P_e\) is typically higher than in mDFT (Fig. 1(b)). DFT-eos uses mDFT for shell-resolved warm-dense electron thermal contributions. From hit-50 ns through the impact window the liner sits at \(\rho\ll\rho_0\) and \(T_e\sim10^2\) eV---partially ionized, warm-dense Al. There \(P_e(\rho,T)\) and \(u_e(\rho,T)\) enter the MHD closure and set the effective adiabat under Joule heating and rarefaction. Along expansion, QEOS tends to maintain lower density in the low- to mid-density range: at hit-50 ns, liner mass-weighted \(\langle\rho\rangle\) is ~48\%--65\% lower than with DFT-eos (Table 2; Fig. 3c). Solid cohesion and Maxwell phase regions in DFT-eos further shift the path from near-ambient solid, yielding different \(u_\mathrm{hit}\) and \((\rho,T)\) profiles at geometric contact (Table 2).

Impact-time contact state makes the path split visible. DFT-eos gives higher \(\rho_\mathrm{in}\) (most clearly at \(\tau=220\) and 260 ns) but lower \(T_{e,\mathrm{in}}\) at \(\tau=260\) and 300 ns (Table 3(a)). Mass-weighted inner \(T_e\) and outer-interface \(T_r\) are not the same quantity. QEOS ends in a hotter, sparser inner state; DFT-eos ends denser and relatively cooler. The two cases strike the low-impedance foam (\(\rho_0=0.015\,\mathrm{g/cm^3}\)) with different \(u_\mathrm{hit}\) and internal-energy flux, so shock strength and the near-interface warm-dense structure differ around \(t_\mathrm{hit}\).

At \(t_\mathrm{hit}\), relative differences in \(S^+\) reach ~50\%--56\% (Table 4)---far larger than the few percent at peak \(S^+\) within the post-impact 15 ns window, where DFT-eos even falls below QEOS at \(\tau=260\) and 300 ns. The contrast reflects how quickly the outer-interface radiation boundary condition forms near impact. Table 3(a) gives \(T_r\) relative differences of ~10\%--12\% at impact; a \(T_r^4\) scaling yields outgoing-power shifts of ~49\%--58\%, matching \(S^+\). At \(t_\mathrm{hit}\) the shock has just entered the foam and the outer surface is still building a radiating boundary layer. Al-side \(P_e\), \(u_e\), and total pressure \(P\) from the chosen EOS enter a Rankine--Hugoniot contact jump that partitions mechanical energy into foam internal energy and outward radiation. TF averaging and mDFT shell structure give different compressibility and energy partition at the same \((\rho,T)\); together with \(u_\mathrm{hit}\) and \(\rho_\mathrm{in}\), this yields a different effective \(T_r\) at the boundary.

Over the next 15 ns, peak relative differences in \(\rho_\mathrm{in}\), \(T_{e,\mathrm{in}}\), \(T_r\), and \(S^+\) shrink as foam heating and multigroup transport average the initial boundary mismatch. Peak \(S^+\) can nearly agree while \(S^+\) at \(t_\mathrm{hit}\) and FWHM still differ (Table 4; Fig. 5c). On this light foam, foam \(E_\mathrm{th}\) and \(\eta\) differ by mostly $\lesssim$10\%, and DFT-eos does not give higher \(\eta\); post-impact foam kinetic energy is comparable to foam \(E_\mathrm{th}\), so bulk motion remains significant. \(\eta\) measures thermalization into foam internal energy; \(T_r\) and \(S^+\) measure the radiation field---the Al electron EOS shows up far more clearly in the latter than in \(\eta\).

Wide-range MHD is selectively sensitive to the Al table. Integrated, path-insensitive quantities (\(t_\mathrm{hit}\), \(E_{k,\mathrm{peak}}\), \(\eta\)) change little; quantities that depend on \(P(\rho,T)\) along the trajectory or on boundary radiation at impact (\(\langle\rho\rangle\), \(\rho_\mathrm{in}\), \(T_r\) and \(S^+\) at \(t_\mathrm{hit}\), FWHM) respond strongly to the electron model---order-50\% \(S^+\) shifts at impact rather than a minor tabulation drift.

\section{Conclusion}

Using the Sec.~III load at 50 MA peak current and rise times of 220--300 ns, we compared DFT-eos with QEOS while holding opacity and electrical conductivity fixed and varying only the Al liner EOS. Impact time and peak liner kinetic energy change little between the two tables, as does conversion efficiency \(\eta\) on this light foam. Differences instead appear in flight-segment density, contact state at impact, and the time structure of outgoing radiation at the outer foam interface---relative differences in \(S^+\) at \(t_\mathrm{hit}\) reach ~50\%, while post-impact peak \(S^+\) falls back to the few-percent level. Opacity and conductivity are the same in both cases, so the radiation discrepancy is driven by the Al electron EOS through the hydrodynamic path: it alters radiation boundary conditions at the foam outer interface, not the absorption coefficient table.

QEOS uses a TF-type model for warm-dense Al electron pressure and internal energy; DFT-eos retains shell structure and partial-ionization physics through mDFT. The two tables give different \(P_e(\rho,T)\) and \(u_e(\rho,T)\), so the liner follows different thermodynamic paths through rarefaction and impact and arrives at contact with different \(\rho_\mathrm{in}\), internal-energy flux, and \(u_\mathrm{hit}\). Post-impact thermalization can bring peak \(S^+\) close even when impact-time \(S^+\) and FWHM still differ (Sec.~V). For dynamic hohlraum applications that depend on impact-time radiation power or pulse shape, wide-range MHD should not rely on TF-type engineering tables alone; shell-resolved Al electron thermodynamics aligned with DFT benchmarks is needed.

\section*{Acknowledgments}
This work was supported by the Science Challenge Project (No.~TZ2025013) and the Funding of National Key Laboratory of Computational Physics (No.~6142A05240101).

\bibliographystyle{apsrev4-2}
\bibliography{references}

\end{document}